# Biclustering Using Modified Matrix Bandwidth Minimization and Biogeography-based Optimization

Briti Deb and Indrajit Mukherjee

**Abstract**

Data matrix having different sets of entities in its rows and columns are known as two mode data or affiliation data. Many practical problems require to find relationships between the two modes by simultaneously clustering the rows and columns, a problem commonly known as biclustering. We propose a novel biclustering algorithm by using matrix reordering approach introduced by Cuthill-McKee's bandwidth minimization algorithm, and adapting it to operate on non-square and non-binary matrices, without the need to know apriori the number of naturally occurring biclusters. This transforms a two-mode matrix into almost block diagonals, where the blocks indicate the clusters between the two modes of the matrix. To optimize the bandwidth minimization problem, we adapted the Biogeography-based Optimization algorithm using logistic equation to model its migration rates. Preliminary studies indicate that this technique can reveal the underlying biclusters in the data and has potential of further research for two-mode data analysis.
Keywords— Biclustering, Bandwidth Minimization, Biogeography-based Optimization, Seriation, Matrix Reordering

## 1 Introduction

Many real world data (also known as affiliation data) have two-mode nature. Given the data is represented as a matrix and the rows (or columns) of the matrix represent actors and columns (or rows) represent groups, a non-zero element in the matrix represents an actor's affiliation to a group. One example of such data is the sales data where the two modes can be the customers and the products/services they consume. Another example is the gene expression data where the two modes can be the genes and the samples on which they express. Yet another example is the co-citation data where the two modes can be the authors and papers they cite. More examples of two-mode data can be found in web query-advertiser network where advertisers (one mode) bids on queries (another mode) and the highest bidder gets its ad placed. For analyzing such two-mode data, it is not sufficient to obtain either row clustering or column clustering using traditional one-mode clustering algorithms. In such cases, simultaneous clustering of columns and rows are necessary which can only be obtained using biclustering.

Most of the traditional techniques that can perform two-mode clustering such as biclustering [1] and Heatmap [2] use some distance metric between the matrix elements. However, such approaches have the drawback of imposing some preconceived structure such as hierarchies, cliques, and structural balance [3]. To circumvent such drawback, we formulate a new biclustering approach by adapting the matrix bandwidth minimization (MBM) algorithm to reorder the columns and rows of the matrix to reveal the two-mode relationships in the data as block diagonals.

The standard definition of matrix bandwidth is the largest distance between any two non-zero elements in any row of the matrix [4]. The MBM problem finds a change in the arrangement (permutation) of the rows (or columns) of a matrix by reordering the rows (or columns) such that the elements which are non-zero in a row (or column) are moved towards the main diagonal [5]. As matrix reordering is a NP complete problem [6] [7], several heuristic algorithms have been developed to address the MBM using row (or column) reordering. One such is the Cuthill-McKee MBM algorithm [5]. However, as the bandwidth minimization problem has a large search space problem, several metaheuristic approaches such as genetic programming [9], Tabu search [10], particle swarm [11], to name a few, have been developed. In this paper, we introduce another approach to address the large search space problem of permuting the rows and columns of the MBM by adapting the Biogeography-based Optimization (BBO) [12] algorithm.

We develop a modified BBO algorithm as an optimizer for the adapted Cuthill-McKee MBM problem. Biogeography [13] which has been the inspiration of BBO, tries to explain how biological species are distributed in space and time. Our reasons for choosing the BBO as an optimizer are: (i) BBO can be used for a wide range of problems as no problem-specific assumptions are made, (ii) its not necessary for the BBO to compute the function gradient, which makes it useful for discrete data, (iii) BBO uses an efficient single fitness evaluation in each generation (it does not evaluate every candidates in each generation, rather only compare the best solution in the previous generation with the new fitness value of the immigrating island), (iv) BBO helps to avoid convergence to local optima due to its features such population, and (v) satisfactory performance of BBO on several benchmarks [14]. Using our modified MBM and BBO algorithm, a binary or non-binary matrix and square or non-square matrix can be transformed into a matrix of almost block diagonals where the blocks indicate the biclusters, without requiring any prior knowledge of the number of biclusters.

There are several applications of biclustering. It can be used to analyze biological data [15], community detection in social networks [16], marketing applications [17], and recommendation systems [17]. Other than finding biclusters in two-mode data, the obtained almost block diagonals can also be used to reduce storage cost for large sparse matrices through dimensionality reduction (also known as subspace clustering) [18].

The main contributions of this paper are summarized as follows: (1) We adapted the Cuthill-McKee MBM algorithm which uses matrix reordering to obtain almost block diagonals of a two-mode matrix where the blocks indicate biclusters. Though the original Cuthill-McKee MBM works for square matrices only, our biclustering algorithm works for both square and non-square matrices. (2) We adapted the BBO algorithm to optimize our adapted MBM algorithm using logistic equation to model the immigration-emigration rates of the BBO. Though the original BBO has been applied to continuous optimization problems, we introduced constraints in our modified BBO which allows it to work for optimization problems involving discrete data.

The rest of the paper is structured as follows. Section 2 presents a literature review of algorithms pertaining to bandwidth minimization, evolutionary optimization, and biclustering. In Section 3, we describe our adapted MBM and adapted BBO algorithms for biclustering. Section 4 reports on the results obtained. Finally, we conclude the paper and outline directions for future work in Section 5.

## 2  LITERATURE REVIEW

An early work on biclustering has been performed by Hartigan in 1972 [19]. Later the variance-based work of Cheng and Church [1] on biological gene expression data made the topic of biclustering more popular. In another work Dhillon et al [20] introduced biclustering techniques for files and words, using information theory and bipartite spectral graph partitioning. Biclustering techniques have also been used by Yang et al. [21][22] to associate customers with products for target marketing and recommendation using the MovieLens dataset. For building recommender systems, biclustering has been used in a collaborative filtering approach using Gibbs sampling on the EachMovie dataset by Hoffman and Puzicha [23] and also by Ungar and Foster [24] who used Expectation-Maximization (EM) algorithm.

Biclustering has similarity with blockmodeling and seriation [38] which has its history back in the 1890s. Searching for structure in the data by means of permuting rows/columns in input-output systems has been explored by [39]. Sparse matrix reordering algorithms for cluster identification has been explored by [25].

Several meta-heuristic algorithms have been used for biclustering, some of which are based on global optimization approaches such as simulated annealing and tabu search [26][27], and some others based on local optimization approaches such as steepest descent. A hybrid of GA with hill-climbing was proposed by [28] to address the BMP. A hybrid of ant colony optimization with hill-climbing was proposed by [29] and particle swarm optimization with hill-climbing by [30] both for BMP. Bandwidth-minimization with genetic programming has been explored by [31] and [32] who applied the algorithm on the Harwell-Boeing sparse matrix collection. Tabu search has been used by [33] to reduce sparse matrix bandwidth.

A distributed framework for data co-clustering with sequential updates has been proposed by [34] who evaluated their results both in Amazon EC2 cloud and local cluster of machines. A parallel heuristic algorithm for bandwidth reduction was discussed by [35].

## 3  METHODOLOGY

### 3.1  Formulating the Modified Bandwidth-minimization Approach for Biclustering

Given a nxn symmetrically structured matrix A=[aij] (where aij not equal to 0 for each i, j), the standard measure of bandwidth of A, β(A), is defined as β(A) = max{|i − j|: aij not equal to 0}. In other words, the bandwidth of a matrix is the maximum distance between two elements in any row of the matrix [25]. The idea of using bandwidth minimization towards biclustering is that by minimizing the bandwidth of a matrix by reordering the rows and columns, the same non-zero elements can be brought as close as possible to the main diagonal such that the resulting dot plot (almost block diagonals) can reveal the biclusters in the data. This is what we mean by biclustering or simultaneous clustering of rows and columns.

We assume a given input matrix M(mxn) whose bandwidth has to be reduced by permuting its rows and columns so that the same elements move closer to the diagonal. The standard definition of bandwidth consider the positional value of the data points without considering their magnitude which makes it suitable only for binary data. In order to make our algorithm suitable for both binary and non-binary data, we modify the standard bandwidth definition to be minimized as follows (Eqn 1):

$$\sum_{i,j}^{row,col} a^2_{ij}(i-j)^2 \qquad (1)$$

where aij corresponds to the entry in the ith row and jth column of the input matrix A. This modified bandwidth (Eqn 1) takes into account the magnitude of the individual entries of the matrix along with their row and column indices. This modified bandwidth function is essentially the objective function for the problem. While minimizing the (i-j)2 component of the objective function pushes the data points towards the main diagonal, the aij2 component of the objective function clusters the same data points. This allows our algorithm to handle non-binary data. The minimization of this objective function is expected to generate the biclusters in the form of block diagonals. As the objective function (Eqn 1) does not impose any restriction on the number of rows and columns of the matrix, it enables our algorithm to handle non-square data matrices, a feature contrary to the standard bandwidth-minimization.

Finding an exact solution to the bandwidth minimization problem is hard as its an NP-complete problem, which calls for using heuristics based techniques. We use a heuristics based on the BBO, hereafter called Modified BBO (MBBO). The motivation for adapting BBO has already been discussed in the previous sections.

## 3.2 Formulating Biogeography-based Optimization for our Modified Bandwidth-Minimization Problem

Biogeography is the study of geographical distribution of species among neighboring islands which can be traced to the work of the nineteenth century naturalists such as Charles Darwin. It aims to explain the migration, origin, and extinction of species among neighboring islands.

The Biogeography algorithm aims to model the distribution of species in the nature's way, which is believed to be a general problem solution technique [12]. This motivates us to apply BBO to choose the best among a set of candidate solutions, given that we have a quantifiable measure of the suitability of the given solution, or in other words, the objective function. In BBO, the parameter habitat suitability index (HSI) [12] represent the suitability of a geographical area as residences for biological species. The HSI is comprised of factors such as diversity of topographic features, rainfall, land area, diversity of vegetation, and temperature. HSIs are characterized by a set of variables called suitability index variables (SIVs) [12]. In a habitat, SIVs are the independent variables and HSIs are the dependent variables. A habitat with high HSI is more static in species distribution because they are already saturated with many species compared to a habitat with low HSI. Therefore, such high HSI habitat is considered to be naturally in equilibrium and a good solution. By good condition, it means low inter-species competition and more biodiversity. Migration happens from high HSI habitat to low HSI habitat which is considered to be an addition of new features from high HSI habitat to the low HSI habitat raising the quality of the low HSI habitat.

We model migration (immigration and emigration) in BBO according to the growth and decay in the Lotka-Volterra model [36]. Suppose that we have an objective function that we want to optimize. Further suppose that we have a population of candidate solutions represented as a vector of integers, where each solution considered as a SIV is represented as an integer in the vector. After assessing the goodness of those candidate solutions, the good solutions (habitats) are expected to have high HSI [12], and therefore less competition between species and more biodiversity, and vice versa for bad solutions. As BBO is a model for geographical distributions of species among neighboring islands, an island with high HSI will tend to share information with islands of low HSI, thus making less habitable islands more habitable, and already more habitable islands much more habitable, controlled through the emigration and immigration rates.

The original BBO method invokes two parameters, namely, the immigration and the emigration rates, which varies linearly in the optimization framework. In this paper, we re-define these two parameters to be nonlinear functions. We use the mutualistic growth-decay relationship given by the Lotka-Volterra model, which has similarity with predator-prey, competition, disease, and mutualism, to define these two functions. This model which basically is a pair of first order nonlinear differential equations aims to model the dynamics of ecological systems in nature. Given x denote the number of prey and y denote the number of predator, the change of x and y with time t is given according to eqn 2.

$$\frac{dx}{dt} = \alpha x - \beta xy$$
$$\frac{dy}{dt} = \gamma y - \delta xy$$
(2)

Here the interaction of the two species are represented by the positive real parameters α, β, γ, and δ, and x and y corresponds to the immigration and emigration rates in our modified BBO approach. The motivation behind the use of Lotka-Volterra model to re-define the BBO parameters is the following: as the basic mutualistic networks remain in equilibrium in a natural way, our hypothesis is that the growth and decay in basic mutualistic networks such as the Lotka-Volterra model can model the ecological dynamics better than the linear migration behavior as used in the original BBO. Based on the population size of the problem, we generate, in offline, the immigration and emigration coefficients, which are later utilized by our biclustering algorithm. The coefficients are obtained by numerically integrating the Lotka-Volterra equations using standard solvers for ordinary differential equations.

For the performance of the BBO operators such as growth, decay, mutation, etc. for row/column permutation of the matrix, a good representation scheme that define the growth-decay (or immigration-emigration) and other BBO operators is necessary. To meet this necessity, a list of interchanges of columns or rows (Eqn 3) has been constructed to represent an island I (i.e., a candidate solution).

$$I = \left(M_1(k_s^1, k_d^1), M_2(k_s^2, k_d^2), \cdots M_m(k_s^m, k_d^m)\right)$$
(3)

where Mi ∈ R,C (R indicates interchange of rows, and C indicates interchange of columns) and kis and kid are the columns/rows to be interchanged. Successively the permutations are applied to obtain the optimal bandwidth-minimized matrix.

### 3.3 Our Biclustering Algorithm

```
1: Input: Data matrix X
2: Output: Data matrix X with almost block diagonals
3: generationLimit E
4: for each e in generationLimit
5: for each mode m in X
6: problemDimensionVector[nrow(X)]= sequence(1:nrow(X))
7: randPermMatrix= permutation(problemDimensionVector)
8: currentEpoch = 0
9: previousBandwidth = ∞
10: bestBandwidth= getBestBandwidth(problemDimensionVector)
11: for each p in problemDimensionVector
12: if (bestBandwidth< previousBandwidth)
13: previousBandwidth= getBestBandwidth(problemDimensionVector)
14: selectIndex = BBO(emigration, immigration)
15: select a row in randPermMatrix based on selectIndex
16: swap the cell value at the pth position of the selected row of Randperm matrix with the cell value at pth position of the problem dimension vector.
17: bestBandwidth= getBestBandwidth(problemDimensionVector)
```

18: end for each p in problemDimensionVector
19: end for each mode m in X
20: end for each e in generationLimit

# 4 RESULTS

## 4.1 Dataset Description

We present the performance of our biclustering algorithm on three datasets: (i) Galaskiewicz's CEOs and Clubs data which contains membership of 26 corporations' chief executive officers in 15 clubs [37], (ii) Southern Women data which contains membership of social events of 18 women who met in a series of 14 informal social events over a period of nine-months [8], and (iii) a synthetic non-binary data partitioned into discrete rectangular chunks. In all these datasets, the actual membership of an entity (row) to a group (column) are already known, hence these datasets also acts as the ground truth for validating the results of our approach. Before running our biclustering algorithm, we randomly permuted the rows and columns of the data, and then ran our algorithm to reconstruct the biclusters. These biclusters are then compared with the ground truth for validation.

## 4.2 Software

The experiments were conducted on an Intel(R) Core(TM) i7 CPU L 640 @2.13 GHz processor computer with 6 GB of RAM running on Windows 8.1 Professional. Matlab and R tools were used to code the program.

## 4.3 Simulation Parameters

Setting the parameters for BBO is a tricky task. We examined the efficacy of different parameters on how many generations it takes to get an acceptable solution. The consistency of the results were checked by running the algorithm for several times. On an average, the execution time taken was 5.5 sec. The datasets used for our experiments have been described in Table I and the visualization of biclusters obtained by applying our algorithm has been shown in Fig. 1.

| Table 1 | | |
|---|---|---|
| Dataset | Data Type | Data Size |
| Galaskiewicz's CEOs and Clubs Data | Binary | 26 x 15 |
| Southern Women Data | Binary | 18 x 14 |
| Synthetic Data | Non-binary | 56 x 50 |

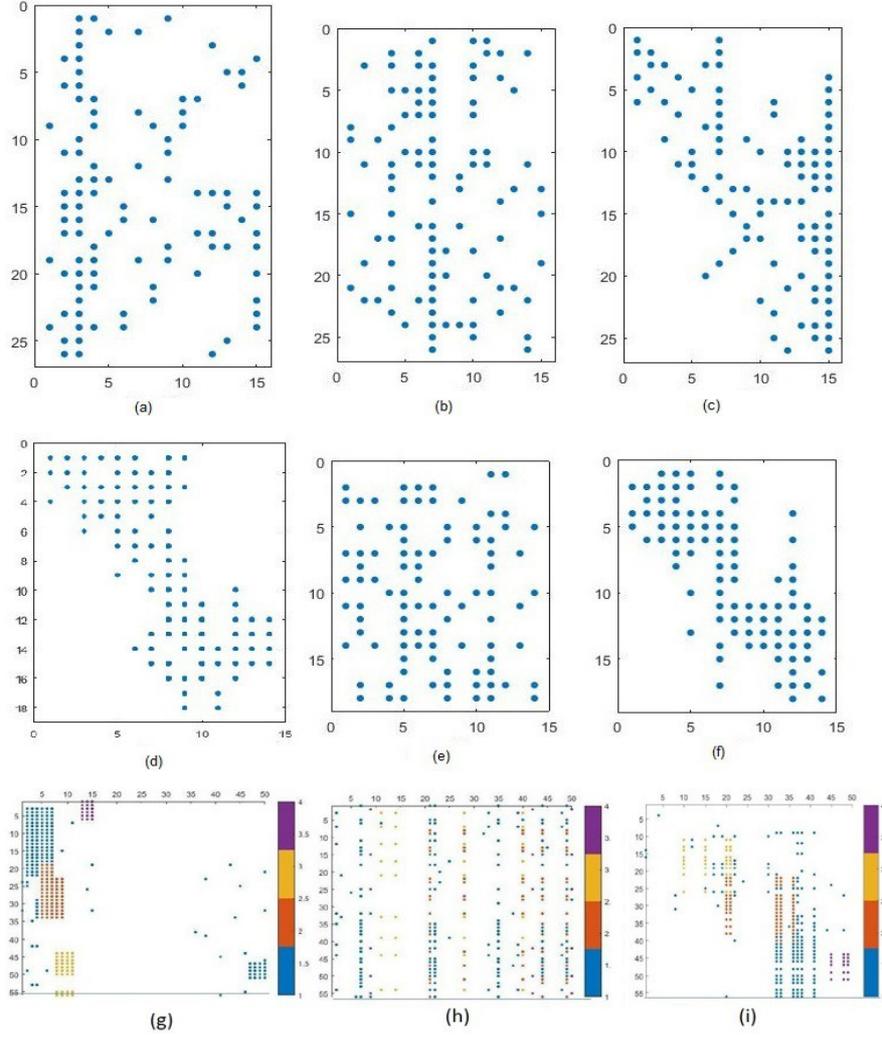

Figure 1: Results showing the biclusters obtained using our approach on the Galaskiewicz's CEOs Clubs data in the first row, Southern Women data in the second row, and our synthetic data in the third row. (a), (d) and (g) are original data which is also the ground truth, (b), (e), and (h) are after random permutation of rows and columns, and (c), (f), and (i) are the obtained biclusters.

## 5    CONCLUSIONS

In this paper we presented the preliminary results of a novel biclustering algorithm based on modified matrix bandwidth minimization and biogeography-based optimization. We tested our modified MBM and BBO-based heuristic with mutualistic growth-decay model on three datasets, and showed that it can extract biclusters satisfactorily. The results indicate that the proposed algorithm provides an alternative approach for biclustering using a meta-heuristic optimization based on BBO. The proposed approach is free from some of the drawbacks of popular biclustering algorithms, such as the requirement to know apriori the total number of clusters, and the use of some distance metric between the matrix elements which impose some pre-conceived structures by the investigator.

In future, we would like to perform experiments on larger datasets. Another area which we would like to work on is improving the evaluation criteria. Last but not the least, we would like to perform probabilistic analysis of the algorithm to prove that the population optimum improves after every generation.